\documentclass[conference]{IEEEtran}
\IEEEoverridecommandlockouts
\usepackage{cite}
\usepackage{amsmath,amssymb,amsfonts}
\usepackage{algorithmic}
\usepackage{graphicx}
\usepackage{textcomp}
\usepackage{xcolor}
\usepackage{comment}
\usepackage{multirow}
\usepackage{array}
\usepackage{listings}
\def\IEEEtitletopspace{20pt}
\newcolumntype{P}[1]{>{\centering\arraybackslash}p{#1}}
\renewcommand*{\arraystretch}{1.0}
\def\BibTeX{{\rm B\kern-.05em{\sc i\kern-.025em b}\kern-.08em
    T\kern-.1667em\lower.7ex\hbox{E}\kern-.125emX}}

\usepackage{tikz}
\usetikzlibrary{shapes}
\usepackage{xcolor}
\newlength\MAX  
\newcommand*\Chart[1]{#1~\rlap{\textcolor{black!20}{\rule{\MAX}{2ex}}}\rule{#1\MAX}{2ex}}
\usepackage{subcaption}
\usepackage{graphicx}
\usepackage{tabularx}
\usepackage{listings}

\begin{document}
 \makeatletter
\newcommand{\linebreakand}{%
\end{@IEEEauthorhalign}
\hfill\mbox{}\par
\mbox{}\hfill\begin{@IEEEauthorhalign}
}
\makeatother
\title{Combining Embedding-Based and Semantic-Based Models for Post-hoc Explanations in Recommender Systems \\
}
\author{
	\IEEEauthorblockN{Ngoc Luyen Le\IEEEauthorrefmark{1}\IEEEauthorrefmark{2}\\\textit{ngoc-luyen.le@hds.utc.fr}}
	\and
	\IEEEauthorblockN{Marie-Hélène Abel\IEEEauthorrefmark{1}\\\textit{marie-helene.abel@hds.utc.fr}}
	\and
	\IEEEauthorblockN{Philippe Gouspillou\IEEEauthorrefmark{2} \\\textit{p.gouspillou@vivocaz.fr} }
	\linebreakand
	\IEEEauthorblockA{
		\IEEEauthorrefmark{1}Université de Technologie de Compiègne, CNRS, Heudiasyc (Heuristics and Diagnosis of Complex Systems),\\ CS 60319 - 60203 Compiègne Cedex, France.
		\\
		\IEEEauthorrefmark{2}Vivocaz, 8 B Rue de la Gare, 02200, Mercin-et-Vaux, France.
	}
}
\maketitle
\begin{abstract}In today’s data-rich environment, recommender systems play a crucial role in decision support systems. They provide to users personalized recommendations and explanations about these recommendations. Embedding-based models, despite their widespread use, often suffer from a lack of interpretability, which can undermine trust and user engagement. This paper presents an approach that combines embedding-based and semantic-based models to generate post-hoc explanations in recommender systems, leveraging ontology-based knowledge graphs to improve interpretability and explainability. By organizing data within a structured framework, ontologies enable the modeling of intricate relationships between entities, which is essential for generating explanations. By combining embedding-based and semantic based models for post-hoc explanations in recommender systems, the framework we defined aims at producing meaningful and easy-to-understand explanations, enhancing user trust and satisfaction, and potentially promoting the adoption of recommender systems across the e-commerce sector.
\end{abstract}


\section{Introduction}
Recommender systems have gained significant prominence in today's information-rich environment, playing a crucial role in helping users navigate the vast array of choices by curating and presenting relevant content through personalized suggestions and recommendations. By analyzing user preferences and item characteristics, these systems efficiently refine the decision-making process, enabling users to make informed choices with confidence. In various domains, such as financial services, luxury goods, real estate, or automobiles, items are often purchased less frequently and tend to be more expensive compared to other commodities. In these cases, the explanation behind why an item is recommended can be as important as the precision of the recommendation itself \cite{zhang2020explainable}.

Embedding-based models are a widely used approach for recommender systems, where user and item feature representations are learned as vectors in a high-dimensional space. These feature representation vectors are then utilized as input for specific matrix factorization or neural network models to predict recommendations \cite{zhang2016collaborative}. However, one major limitation of embedding-based models is their lack of interpretability, which can result in diminished trust and user engagement \cite{wang2019explainable}. Furthermore, the integration of ontology-based knowledge graphs in recommender systems as semantic-based models has proven to be an effective approach, generating reliable recommendations while also enhancing interpretability and explainability \cite{palumbo2020entity2rec, zhang2020explainable}. 
 Generally, explanations can be generated using two main strategies: (i) the adoption of transparent models that provide explanations during the recommendation generation process, and (ii) post-hoc explanations, which provide justifications for the recommendations after they have been generated, even when the underlying model is complex or not directly interpretable \cite{zhong2022shap}.

Despite their widespread adoption and success in providing accurate recommendations, many recommender systems operate as ``black boxes", offering little insight into the rationale behind their suggestions. This causes the lack of transparency that can result in reduced user trust and satisfaction, and acceptance of the recommendations provided \cite{bilgic2005explaining, saeed2023explainable,alshammari2019mining }. Recent advances in embedding-based and semantic-based models offer promising approaches to enhance recommendation precision and explanation by capturing complex patterns and relationships in data or leveraging structured knowledge and interpretable features \cite{chen2019collaborative, confalonieri2021using, alshammari2019mining}. However, the potential for integrating these models to provide post-hoc explainability has not been fully  exploited. In this paper, we explore an approach that combines embedding-based and semantic-based models for post-hoc explanations in recommender systems. By combining the complementary strengths of these approaches, we defined a framework to generate meaningful and interpretable explanations, enhancing user understanding of the rationale behind item recommendations. 

The rest of this article is structured as follows: Section \ref{sec_relatedworks} provides an introduction to the relevant literature that underpins our approach. In Section \ref{sec_our_approach}, we present our primary contributions, which detail how we combine embedding-based and semantic-based models to construct post-hoc explanations in recommender systems. Section \ref{sec_experiment_and_evaluation} showcases our experimentation with a dataset in the domain of buying and selling used vehicles, where we apply our approach. Lastly, we conclude by discussing the implications and future prospects.
\section{Related work}\label{sec_relatedworks}
\subsection{Embedding-based recommender system}
Embedding-based recommender system (RS) uses embedding feature representation vectors to represent items and users in a high-dimensional space. Therefore, items are similar if their characteristics or features have similar embedding and users have similar preferences if they also have similar embedding vectors. The process of generating embedding vectors  are employed by training a neural networks on a large dataset of item user interaction or a knowledge bases related to user and item \cite{dong2021hybrid}.  

In the past couple of years, many recent studies have shown that embedding-based models can outperform traditional recommendation algorithms. For example, Liu et al. \cite{liu2020hybrid} proposed a hybrid recommendation model that combines collaborative filtering and content-based approaches using embedding-based representations. Their proposed model achieved state-of-the-art performance on several benchmark datasets. Pan et al. \cite{pan2020explainable} proposed the use of embedding vectors by mapping general features learned using a base latent factor model onto interpretable aspect features. This approach allows for a trade-off between explainability and performance in their RS. Palumbo et al. \cite{palumbo2020entity2rec} proposed constructing knowledge embeddings for the RS by learning feature representations based on the properties of user and item entities. 
 However, the use of embeddings can make it difficult to interpret the reasons behind the item recommendations generated by the model \cite{he2017neural}. Furthermore, these models may not always capture the full range of user preferences or contextual information that can be important factors in predicting recommendations \cite{wang2019kgat}. In the next section, we will explore an approach that leverages ontology-based knowledge graphs to address these limitations and enhance the explainability of RSs.
\subsection{Semantic-based recommender system}
The use of ontology-based knowledge graphs as semantic-based models in RSs has emerged as a promising approach to address some of the limitations of embedding-based models. Ontology-based knowledge graphs can capture rich contextual and semantic information, as well as relationships between items and users \cite{arafeh2021ontology}. These characteristics can be used to develop feature learning techniques that can enhance both the precision and explainability of RSs \cite{guo2020survey,zhang2020distilling}. For example, Ishita et al. \cite{padhiar2021semantic} proposed an ontology that provides a formalism for modeling explanations to users and for item recommendations in the food domain. Kamma et al. \cite{kamma2021recommendation} proposed to generate product recommendations and explanations using ontological reasoning that finds and infers knowledge paths in ontology models. Le et al. \cite{le2022towards} proposed the development of an ontology in the vehicle domain and its use in generating relevant item recommendations transparently. However, there are also some disadvantages to using ontology-based knowledge graphs in RSs. One challenge is the scalability of the approach as engineering and maintaining a specific domain ontology can be time-consuming and resource-intensive \cite{simperl2006ontology, luyen2016development}. Moreover, there may be difficulties in integrating the ontology-based knowledge graph with existing recommendation algorithms or systems \cite{tiddi2022knowledge}.
By making the relationships between items and users explicit and providing a formal representation of the knowledge and concepts underlying the recommendations, semantic-based models can generate transparent and interpretable item recommendations. The combination of embedding-based and semantic-based models can benefit the precision and explainability of RSs. In the section, we will explore about post-hoc explanations in RSs.

\subsection{Post-hoc explanations}
Post-hoc explanations in RSs refer to the generation of explanations after the recommendation has been provided. These explanations can be presented to the user in order to provide insights into why a particular item was recommended, and can help to increase user trust and satisfaction with the system \cite{zhong2022shap}. In general, post-hoc explanation approaches are categorized into the following methods: rule-based methods, feature importance methods, and counterfactual explanations. Firstly, rule-based methods use explicit rules to explain the recommendations, 
for example, Peake et al. \cite{peake2018explanation} developed an association rule mining approach based on user history as input to explain the outputs of a matrix factorization recommendation model. Then, counterfactual explanation methods involve generating alternative recommendations and explaining why they were not chosen. 
For instance, Wang et al. \cite{wang2022reinforced} proposed a counterfactual explanation approach for item recommendations based on item attribute-based models. 
Lastly, feature importance methods identify the features that were most important in generating the recommendation, such as the user's past purchase history or ratings. For example, Ribeiro et al. \cite{ribeiro2016should} proposed LIME (Local Interpretable Model-agnostic Explanation) as an example, which utilizes sparse linear models to approximate a complex and non-linear classifier around a sample. 
Furthermore, Lipton \cite{lipton2018mythos} proposed the SHAP (SHapley Additive exPlanations) method based on Shapley values, a concept derived from cooperative game theory. The SHAP method calculates the contribution of each feature to a prediction by considering all possible coalitions of features. 

Understanding the importance of the features returned by LIME and SHAP is not trivial as it requires complex background mathematical knowledge, such as knowledge in game theory. LIME or SHAP can be a potential approach for model developers to visualize the behaviors of models and help them check their models. Nonetheless, the explanations of these approaches may not be easily understandable for end-users in e-commerce applications \cite{zhong2022shap}. The integration of semantic-based models with post-hoc explanations in RSs has the potential to enhance the transparency and interpretability of the generated recommendations. Therefore, in the following sections, we will explore how the combination of embedding-based and semantic-based models can be leveraged for post-hoc explanations in RSs.

\section{Our Approach}\label{sec_our_approach}
This section will introduce our approach for post-hoc explanation in a recommender system, which not only aims to recommend items to users but also to offer explanations for the recommended items. To demonstrate this idea, we will utilize a vehicle knowledge graph represented through ontologies from the e-commerce domain, which specifically pertain to the buying and selling of vehicles.

\subsection{Task Formulation}
In the domain of e-commerce applications, the data gathered for the construction of a typical recommender system can be decomposed into three essential constituents: a set of items, denoted by $I = \{i_1, i_2, .., i_n\}$, a set of users, represented by $U = \{u_1, u_2, ..., u_3\}$, and a set of interactions between users and items, symbolized by $D = \{d_{ui}| u \in U, i \in I\}$. In this formulation, $d_{ui} = 1$ is an indicator that a user-item interaction has occurred between user $u$ and item $i$, while $d_{ui}=0$ indicates that no such interaction has taken place. In addition, an ontology-based knowledge graph can be integrated to contribute more explicit and implicit information about users and items. This knowledge graph is represented as a set $G = (E, R)$, where $E$ denotes the set of entities, and $R$ represents the set of relations between these entities. In the context of a recommender system, an entity $e \in E$ of the knowledge graph $G$ can be a user $u \in U$ or an item $i \in I$. A relation $r \in R$ serves as a predicate in a triple $\langle subject, predicate, object \rangle$, representing the relationship between two entities or the relation between an entity and its property.

Given a collection of interactions between users and items in an e-commerce application and the ontology-based knowledge graph $G$, the recommendation task within the scope of this work is divided into two separate tasks. The first task involves predicting a list of the most suitable items for a specific user. The second task focuses on generating explanations for the recommended items, taking into account the particular context and user preferences. The explanation task emphasizes providing instance-level explanations to justify the relevance of each recommendation to the user.

\begin{figure}[htbp]
	\centerline{\includegraphics[width=0.7\linewidth]{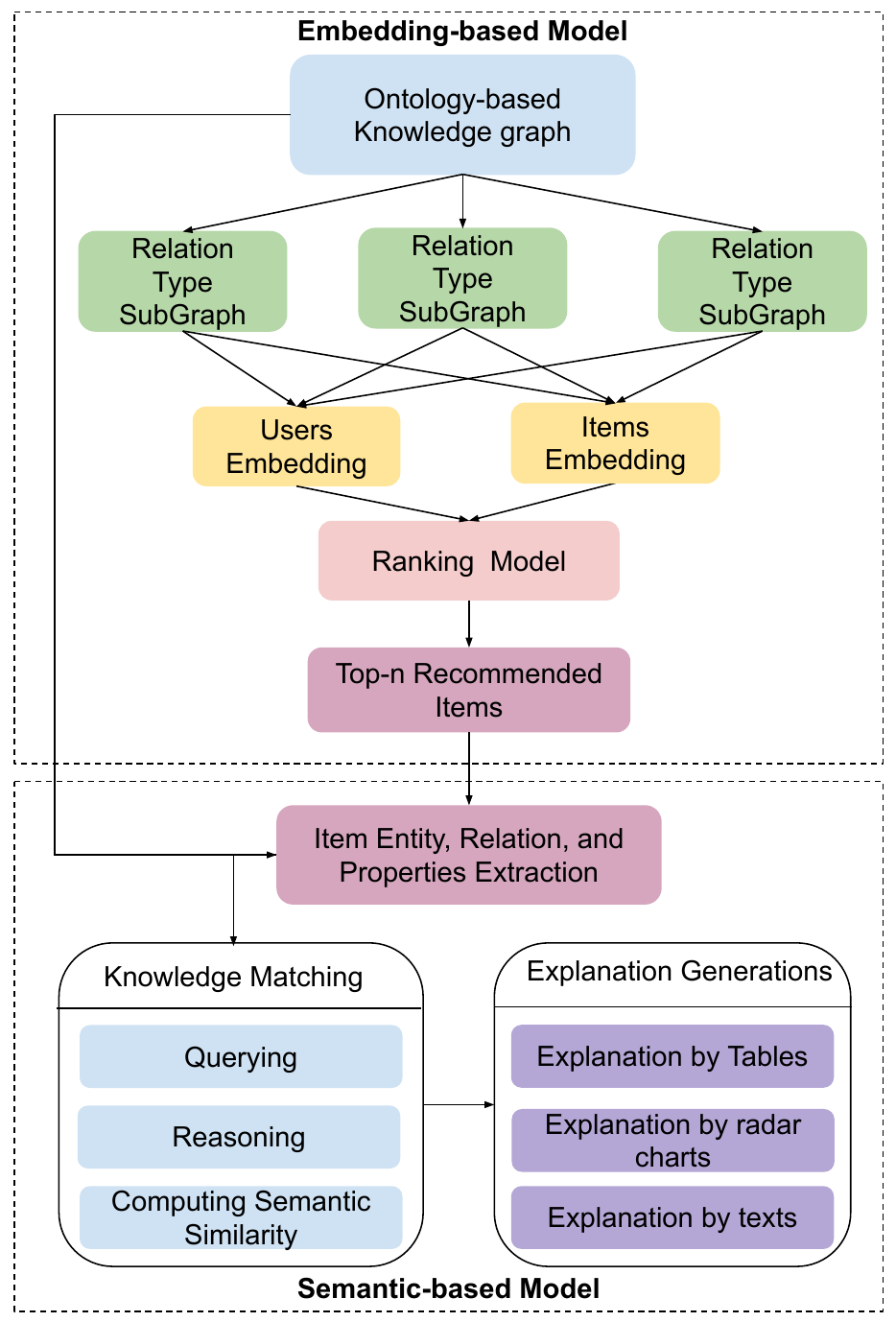}}
	\caption{Schematic diagram of the approach for building a post-hoc explanation in a recommender system, employing embedding-based and semantic-based models.}
	\label{fig_01}
	\vspace{-0.5cm}
\end{figure}

\subsection{Our Framework Architecture}
In this section, we present the architecture of our framework, which aims to address the challenges of integrating and combining embedding-based and semantic-based models for post-hoc explanations in RSs.  As illustrated in Figure \ref{fig_01}, the general architecture of the framework comprises two main components, each corresponding to one of the primary tasks. the generation of top-n recommendations task is accomplished through the utilization of embedding-based models that leverage the ontology-based knowledge graph. The generation of top-n recommendations is achieved using an embedding-based model. The construction of feature representation vectors for items and users is carried out by extracting their information and relationships within the knowledge graph, focusing on subgraphs of specific relation types. Embedding vectors for users and items are learned using knowledge graph embedding methods such as TransE \cite{bordes2013translating}, TransH \cite{wang2014knowledge}, or Entity2rec \cite{palumbo2020entity2rec}. The top-n recommendations are then derived from a learning-to-rank model.

Meanwhile, the post-hoc explanation task is tackled using semantic-based models, which are capable of providing transparent explanations for individual recommendations. To achieve this, the system delves into the details of the recommended item and extracts relevant graph information from the ontology-based knowledge graph through querying and reasoning. By computing the similarity between properties or RDF subgraphs, the framework can synthesize instance-level explanations for the recommended items. These explanation results are subsequently employed to evaluate aspects such as customer satisfaction and the precision of the explanations provided.
The main advantage of embedding-based models lies in their performance and precision, given the available dataset. In the following section, we delve into the details of our embedding-based model and outline the process for generating top-n recommendations.

\subsection{Top-n recommendations based on the knowledge graph embedding-based model}
 This section provide a comprehensive understanding of the knowledge graph embedding-based model's inner workings, showcasing how it effectively leverages an ontology-based knowledge graph and the user-item interaction dataset to produce accurate and relevant recommendations.
 Taking into account various relation types and properties for constructing feature representation vectors, we utilize the widely adopted \textit{Entity2rec} method on the knowledge graph, applied to each relation type, to embed entities into the space $\mathbb{R}^d$, drawing inspiration from the work of \cite{palumbo2020entity2rec}. Initially, we extract and construct a sub-graph for each relation type $G_p$. Subsequently, we learn the vector representation of nodes for each relation type $x_p: e \in G_p \rightarrow \mathbb{R}^d$ . The model is trained using the \textit{node2vec} objective function, as described by Grover et al. \cite{grover2016node2vec}:
 \begin{equation}
 	\max_{x_p} \sum_{e \in G_p}(-log Z_e + \sum_{n_i\in N(e)} x_p(n_i)\times x_p(e))
 \end{equation}
 where $Z_e = \sum_{v\in G_p }exp(x_p(e)\times x_p(v))$ represents the per-node partition function. We approximate it utilizing negative sampling, as proposed by Mikolov et al. \cite{mikolov2013distributed}. The neighborhood of an entity $e$, denoted as $N(e)$, is established using a \textit{node2vec} random walk. The optimization procedure is executed through stochastic gradient ascent on the parameters defining $x_p$.
 For a specific user $u \in U$, top-n item recommendations are obtained from the n-highest scores among the list of candidate items. These scores are derived from a ranking function $\mathfrak{r}(u,i)$ that evaluates the relevance of an item to a user. We employ the LambdaMart model \cite{burges2010ranknet} to obtain these scores. 
 Having obtained the top-n recommended items for a user from the knowledge graph embedding-based models, which function as black box models without providing insights into the item explanations, we now turn our attention to explaining items in accordance with user preferences in the next section. 
 
\subsection{Post-hoc Explanation based on the Semantic-based model}
With a comprehensive understanding of the embedding-based model for generating top-n recommendations, this section explores the methods and techniques employed to produce transparent, instance-level explanations for individual recommendations.
Leveraging the ontology-based knowledge graph $G$, we first extract information pertinent to a user $u$ in the form of a subgraph RDF. This step allows us to gain valuable insights into the user's preferences and characteristics. Subsequently, for each recommended item $i$ generated by the prediction task, we retrieve its relations and properties within the knowledge graph $G$ using SPARQL queries. This process results in the formation of two distinct subgraphs - one representing user $u$ and the other representing item $i$. We then identify pairs of properties shared between the user's preferences and the item's description. Drawing on the research by Le et al. \cite{luyen2023, lengochal03675591}, we can calculate the similarity between triples as follows:
\begin{equation}
	Sim(a_{s1},a_{s2}) = \frac{\sum_{i=1}^{k} \bar{S}(w_{1i}, a_{s2}) + \sum_{j=1}^{l} \bar{S}(w_{2j}, a_{s1})}{k + l}
\end{equation} where  $a_{s1}$ and $a_{s2}$ are triples which have word vectors are $M_1 = \{w_{11}, w_{12}, ..., w_{1k}\}$ and $ M_2 = \{w_{21}, w_{22}, ..., w_{2l}\}$,  and $\bar{S}(w, a_{s})$ denotes the semantic similarity of a word $w$ and a triple. The function $\bar{S}(w, a_{s})$ is formally calculated as follows:
	$\bar{S}(w, a_{s}) = \max\limits_{w_i \in M} \bar{S}(w, w_i)$
 where $w_i \in M=\{w_1, w_2, ..., w_k\}$ is the word vector of $a_s$. Each word $w_i$ is represented by a numerical vector. Finally, we can establish the percentage of matching between user $u$ and item $i$ by computing the semantic similarity between their subgraphs RDF.

We demonstrate the process of generating explanations using an ontology developed in the domain of purchases and sales, as presented by the authors in \cite{le2022towards}. Consequently, we enrich the ontology-based knowledge graph by feeding it with instances of users and items, as an example illustrated in Figure \ref{fig_02}. Apart from retrieving subgraph RDFs of items or users using SPARQL queries, we can also deduce implicit information by using rules. For instance, if a user is interested in environmentally friendly vehicles, the vehicle model ``Tesla Model 3" may appear in their list of favorite vehicles. This is because the ontology specifies that electric cars belong to the subclass of eco-friendly vehicles, and ``Tesla Model 3" is an instance of the electric cars class.
\begin{figure}[htbp]
	\centerline{\includegraphics[width=0.78\linewidth]{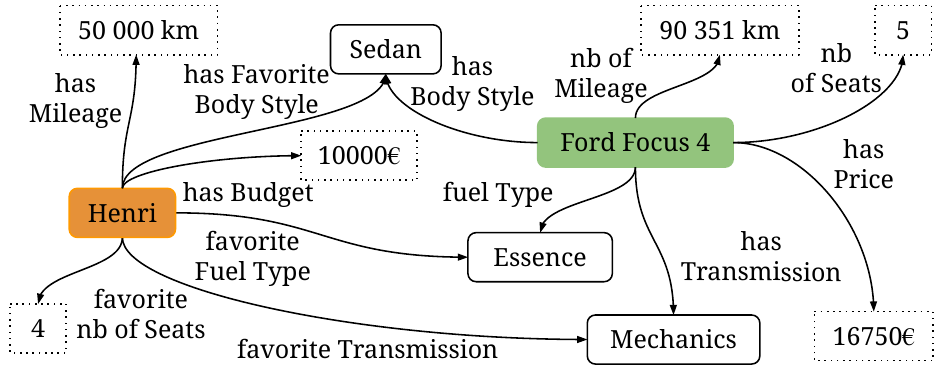}}
	\caption{Snapshot of a section related to instances of users and items - vehicles within the ontology-based knowledge graph}
	\label{fig_02}
	\vspace{-0.2cm}
\end{figure}

Depending on the context and user settings, explanations can be generated in various formats, such as a comparison table, a radar chart, or a textual description. While these presentation forms may differ, they all fundamentally ensure that the explanations convey the same information regarding the relatedness of user preferences and recommended items. This adaptability in presentation style caters to different user needs and preferences, ensuring that the explanations are easily comprehensible and effectively communicate the rationale behind the recommendations.

\section{Experiment and Evaluation}\label{sec_experiment_and_evaluation}
\subsection{Recommendation generation}
We conducted our experiment utilizing a dataset from the vehicle purchase and sale domain, in which users represent clients and items correspond to various vehicle models available in the market. Our dataset consists of 393 users and 5,537 vehicles. The key statistics of this dataset are presented in Table \ref{tab1}.
\begin{table}[h]
	\centering
	\begin{tabular}{| p{2.8cm} | p{2.5cm} | m{1.5cm} |}
		\hline
		\multirow{3}{*}{User - Item Interaction} & Users & 393 \\\cline{2-3}
		& Items & 5.537 \\\cline{2-3}
		& Iteractions & 99.121 \\\cline{1-3}
		\multirow{3}{*}{Knowledge Graph} & Entities & 27.561 \\\cline{2-3}
		& Relation types & 6 \\\cline{1-3}
	\end{tabular}
	\caption{Some statistics on the dataset}
	\label{tab1}	
	\vspace{-0.3cm}
\end{table}

\begin{table}[h!]
	\centering
	\begin{tabular}{| l | P{0.8cm} | P{0.8cm} | P{0.8cm} | P{0.8cm} | P{0.8cm} |}
		\hline
		Approach & P@5 & P@10 & MAP & R@5 & R@10 \\\hline
		TransE & 0.4992 & 0.4625& 0.5690 &0.2652& 0.4034\\\hline
		TransR & 0.4936  & \textbf{0.4653} & \textbf{0.5734} & 0.2768 & \textbf{0.4094}  \\\hline
		WRMF & 0.3195 & 0.3053 & 0.2874 & 0.0793 & 0.1459 \\\hline
		BPRMF & 0.3226 & 0.3076 & 0.3024 & 0.1098  & 0.1926 \\\hline
		ItemKNN & 0.3974 & 0.3809 & 0.4311 & \textbf{0.4113} & 0.1655 \\\hline
		MulticoreBPRMF & 0.3389 & 0.3104 & 0.2987 & 0.0875 & 0.1560 \\\hline
		Our Model & \textbf{0.5002} & 0.4608 & 0.5619 & 0.2805 & 0.4039\\\hline
	\end{tabular}
	\caption{Results achieved using various approaches on the same dataset}
	\label{tab3}
	\vspace{-0.5cm}
\end{table}

To evaluate the recommendation task, we employ standard information retrieval metrics such as Precision at K (P@K), Recall at K (R@K), and Mean Average Precision (MAP). We compare our approach with six other methods: TransE, TransR, WRMF, BPRMF, ItemKNN, and MulticoreBRPMF \cite{Gantner2011MyMediaLite}. Our experiments focus on a feature set comprised of six primary features: \{\textit{Price}, \textit{Transmission}, \textit{Body Style}, \textit{Fuel Type}, \textit{Mileage}, \textit{Number of Seat}s\}. The experimental results, as illustrated in Table \ref{tab3}, indicate that our recommendation approach demonstrates strong performance when compared with the other methods on the vehicle dataset. Our model achieved the best results at the metric of Precision at k=5. 
\begin{figure}[htbp]
	\vspace{-0.8cm}
	\centerline{\includegraphics[width=0.90\linewidth]{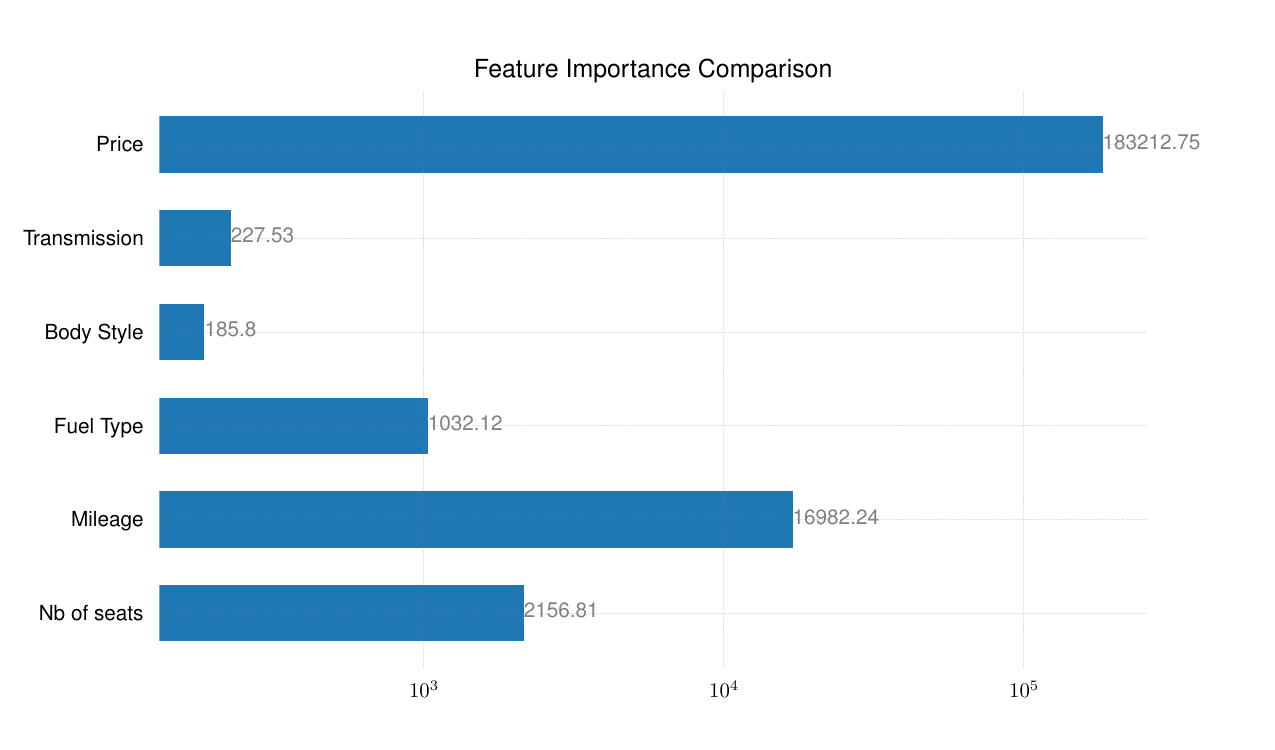}}
	\vspace{-0.5cm}
	\caption{Feature importance derived from the training process}
	\label{fig_03}
	\vspace{-0.3cm}
\end{figure}

Leveraging relation types between entities or between entities and their properties represented in an ontology-based knowledge graph and a semantic-based model, we can thoroughly explore the rich information of users and items along with their relationships. The use of embedding-based models allows for the recommendation tasks to achieve outstanding performance and precision. However, explaining and establishing a connection to demonstrate how the results were obtained remains challenging. For instance, Figure \ref{fig_03} illustrates the importance of features within the training model for the recommendation task. This information assists model developers in modifying and adapting the most suitable features for their model. However, end-users require a better approach to explain the results and understand how they relate to their information, such as historical activities and preferences. 

\subsection{Explanation generation}
Building upon the recommended items generated, the post-hoc explanation step synthesizes and presents explanations for the items in a relevant manner. To illustrate the explanation process, we utilize a simple example depicted in Figure \ref{fig_02}, showcasing three ways explanations can be generated and displayed to end-users. Firstly, explanations can be conveyed through a radar chart, in which each corner symbolizes a feature, and the chart visually communicates the degree of matching between each feature and the user's preferences, as shown in Figure \ref{fig_04}. This visualization allows users to rapidly comprehend the significance of each feature and its contribution to the recommendation. Secondly, a tabular format can be utilized to explicate the features and their alignment with the end-user's needs as shown in Figure \ref{tab_04}. This approach offers a clear and organized presentation of the information, enabling users to easily compare and contrast the features and their respective matching scores. Lastly, explanations can be represented using natural language, which provides a more intuitive and user-friendly approach to conveying the relationship between features and user preferences as illustrated in Figure \ref{fig_05}. 
By presenting information in a conversational manner, end-users can easily understand the rationale behind the recommendations without needing to decipher complex visualizations or tables.

\begin{figure}
	
	\renewcommand*{\arraystretch}{1.4}
	\centering
	\begin{subfigure}{0.24\textwidth}
	\centerline{\includegraphics[width=\linewidth]{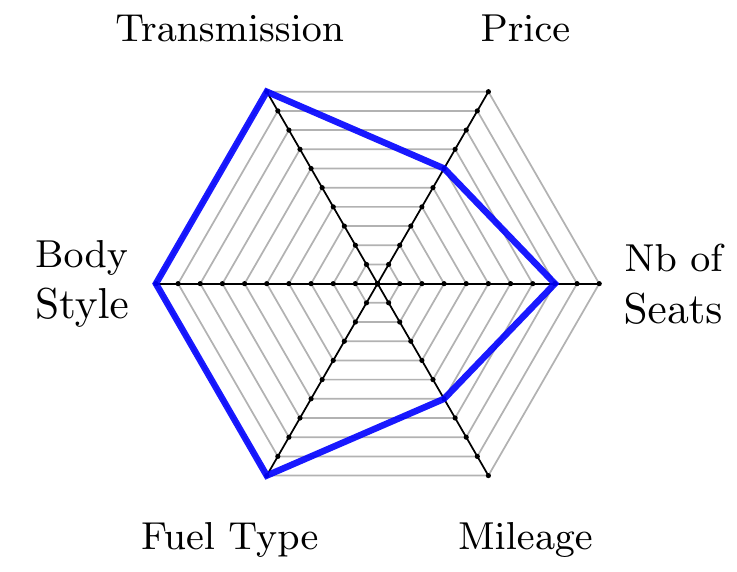}}
	\caption{Explanations using radar chart}
\label{fig_04}
	\end{subfigure}
	\hfill
	\begin{subfigure}{0.24\textwidth}
			\centering
			\resizebox{\textwidth}{!}{\begin{tabular}{| p{2.8cm} | p{2.5cm} |}
				\hline
				User preference & Matching \\\hline
						Price & \Chart{0.66} \\\hline
						Transmission & \Chart{1.00} \\\hline
						Body Style &  \Chart{1.00}\\\hline
						Fuel Type & \Chart{1.00} \\\hline
						Mileage & \Chart{0.66} \\\hline
						Nb of Seats & \Chart{0.83} \\\hline
						Global score & \Chart{0.86} \\\hline
			\end{tabular}
			\vspace{0.2cm}	}
		\caption{Explanations in tabular format}
		\label{tab_04}
	\end{subfigure}
	\hfill
	\begin{subfigure}{0.48\textwidth}
		\centering
		\resizebox{\textwidth}{!}{\begin{tabularx}{\linewidth}{XX}
				\hline {\small
			\textit{The recommended item closely matches your preferences in several key aspects. It has a 66 \% match with your proposed mileage. The vehicle body style, fuel type, and transmission align perfectly with your preferences, with a 100 \% match for both categories. Additionally, the number of seats has an 83 \% match, and the price aligns with your budget at a 66 \% match. Overall, the item demonstrates an 86 \% global match with your preferences, making it a highly suitable choice based on your requirements.}}\\\hline
		\end{tabularx}}
		\caption{Explanations using natural language}
		\label{fig_05}
	\end{subfigure}
	
	\caption{Three different formats for explanations to end-users}
	\label{fig:figures}
	\vspace{-0.5cm}
\end{figure}
\section{Conclusion and Perspective}\label{sec_conclusion}
In this paper, we have presented an approach for generating recommendations and explanations in recommender systems by combining embedding-based and semantic-based models. We capitalize on the advantages of our embedding-based model for generating recommendations, which exhibits high performance and precision compared to other approaches, while employing our semantic-based model with ontology-based knowledge graphs to transparently explore and deliver meaningful explanations to end-users. Our framework, rooted in post-hoc explanation methods, illustrates its effectiveness in the vehicle purchase and sale domain, offering various explanation presentation formats, such as radar charts, tables, and natural language descriptions, to adapt to diverse user needs and preferences. The approach aims to enhance user trust and satisfaction by promoting transparency and interpretability in the recommendation process. In our future work, we plan to carry out public testing of our system and its explanations, with the goal of gathering valuable feedback and insights to guide further enhancements. Moreover, we intend to integrate advanced machine learning techniques and assess the impact of various explanation modalities on user satisfaction.
\section*{Acknowledgment}
This work was funded by the French Research Agency (ANR) and by the company Vivocaz under the project France Relance - preservation of R\&D employment (ANR-21-PRRD-0072-01).
\bibliographystyle{ieeetr}
\bibliography{references}

\end{document}